\documentclass[journal=jpclcd,dvilaser,manuscript=letter,layout=traditional]{achemso}
\usepackage[T1]{fontenc}
\usepackage[latin9]{inputenc}
\usepackage{xcolor}
\usepackage{amsthm}
\usepackage{amsmath}
\usepackage{amssymb}
\usepackage{graphicx}
\usepackage{esint}
\PassOptionsToPackage{normalem}{ulem}
\usepackage{ulem}

\makeatletter

\title{Clustered Geometries Exploiting Quantum Coherence Effects for Efficient
Energy Transfer in Light Harvesting}
\author{Qing Ai}
\affiliation{Department of Chemistry and Center for Quantum Science and Engineering,
National Taiwan University, Taipei City 106, Taiwan}
\author{Tzu-Chi Yen}
\author{Bih-Yaw Jin}
\author{Yuan-Chung Cheng}
\email{yuanchung@ntu.edu.tw}
\providecolor{lyxadded}{rgb}{0,0,1}
\providecolor{lyxdeleted}{rgb}{1,0,0}

\newcommand{\lyxdeleted}[3]{{\color{lyxdeleted}\sout{#3}}}

\usepackage{amsfonts}
\usepackage{bbm}
\usepackage{epsfig}

\usepackage{achemso}

\usepackage{color}

\makeatother

\begin{document}
\begin{abstract}
Elucidating quantum coherence effects and geometrical factors for
efficient energy transfer in photosynthesis has the potential to uncover
non-classical design principles for advanced organic materials. We
study energy transfer in a linear light-harvesting model to reveal
that dimerized geometries with strong electronic coherences within
donor and acceptor pairs exhibit significantly improved efficiency,
which is in marked contrast to predictions of the classical Förster
theory. We reveal that energy tuning due to coherent delocalization
of photoexcitations is mainly responsible for the efficiency optimization.
This coherence-assisted energy-tuning mechanism also explains the
energetics and chlorophyll arrangements in the widely-studied Fenna-Matthews-Olson
complex. We argue that a clustered network with rapid energy relaxation
among donors and resonant energy transfer from donor to acceptor states
provides a basic formula for constructing efficient light-harvesting
systems, and the general principles revealed here can be generalized
to larger systems and benefit future innovation of efficient molecular
light-harvesting materials.

\newpage{}
\end{abstract}
\maketitle
\textbf{TOC Graphic.}

\includegraphics[bb=0bp 0bp 309bp 309bp,width=5cm,height=5cm]{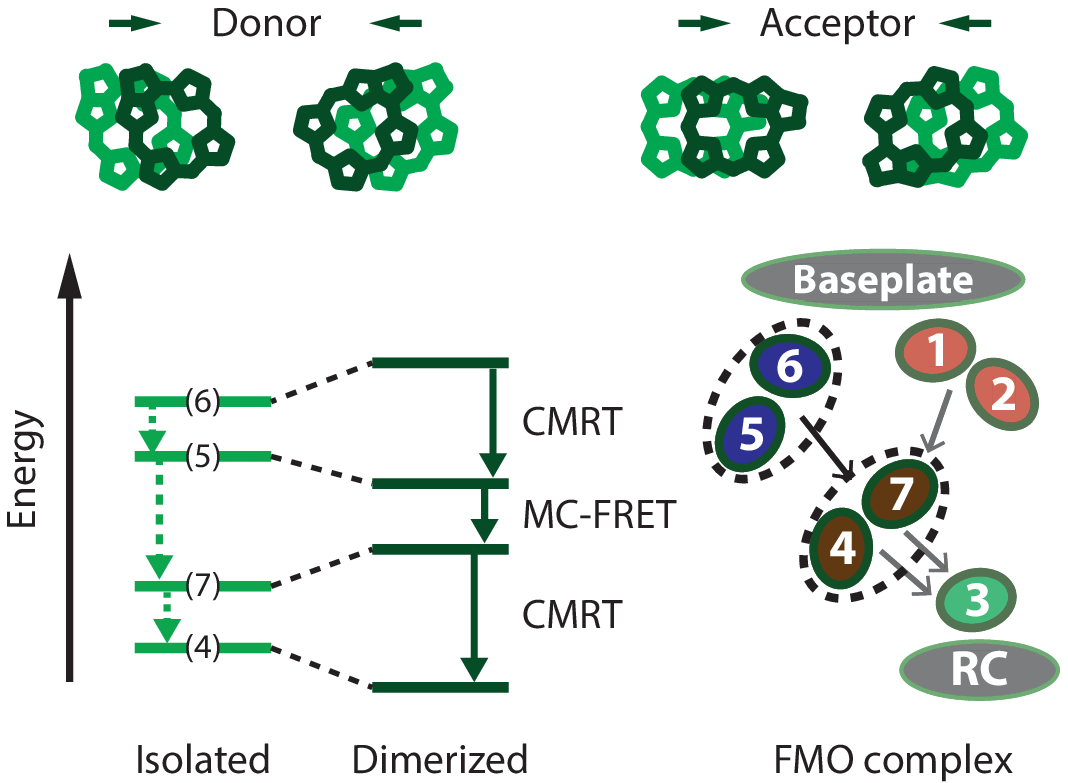}


\newpage{}

Photoactive molecular architectures have recently attracted intensive
research interests in Chemistry and Nanoscience because of their potential
to achieve extraordinary optoelectronic properties. For example, in
natural photosynthesis, sophisticated pigment-protein complexes are
responsible for absorbing light energy and transferring the excitation
energy across tens of nanometers to the reaction center with near
unity quantum efficiency.\cite{Scholes:2006p16773,Cogdell:2008p59896,Cheng:2009p75665,Scholes:2011kx,Lambert:2013fj}
Inspired by these remarkable natural assemblies, artificial molecular
systems have been synthesized to significantly improve the energy-transfer
performance of organic materials.\cite{Huang:2011p93063,Liu:2010p95490}
Despite recent advances, the rational design of such photoactive molecular
materials is still very much under development, because a fundamental
understanding of the structure-function relationships regarding excitation
energy transfer (EET) and charge transfer dynamics in these complex
molecular systems in the condensed phase is still lacking.

Recently, quantum-beat-like phenomena observed in the two-dimensional
electronic spectra of photosynthetic complexes\cite{Engel:2007hb,Collini:2010p79587,Panitchayangkoon:2010p82449}
have inspired numerous research interests to understand the roles
of quantum coherence in natural photosynthesis.\cite{Ishizaki:2009p77158,Mohseni:2008p67347,Chin:2010p81675,Ishizaki:2012kf,Pachon:2012fm,Cao:2009p78157}
While the significance of coherent electronic dynamics in photosynthesis
is still under debate,\cite{Scholes:2010p80927,Pachon:2012fm,Fassioli:2012gd,Kassal:2013bo}
it is clear that quantum effects are important and recent works have
provided much insights into the quantum mechanical control of energy
transport,\cite{Rebentrost:2009kv,Ishizaki:2009p77158,Cao:2009p78157,Chin:2010p81675,Wu:2010bg,Ishizaki:2012kf,Kassal:2013bo,Jang:2012fi}
including the significance of initial preparation.\cite{Moix:2011gi,Wu:2013ig}
We believe now it is the time to go beyond detection and to confront
the more relevant question in real-world applications: how to utilize
electronic quantum coherence to build a more efficient EET network?

Before we go on to discuss coherence effects in EET, we believe a
more precise definition of coherence in excitonic systems is necessary.
The term ``coherence'' generally refers to non-zero off-diagonal
elements of the density matrix describing a quantum system. Since
the values of a density matrix are basis-dependent, quantum coherence
in EET phenomena could refer to different concepts.\cite{Cheng:2009p75665,Pachon:2012fm}
In the site basis, in which electronic excitations localized on individual
chromophores are used as the basis states, coherence represents coherent
delocalization of excited states. In contrast, in the exciton basis,
which are delocalized eigenstates of the electronic Hamiltonian, coherence
represents superposition of eigenstates, hence a non-stationary state
that evolves coherently under the electronic Hamiltonian.\lyxdeleted{Yuan-Chung Cheng}{Wed Jul 17 17:47:55 2013}{
} It is clear that exciton delocalization plays an important role
in the efficient harvesting of light energy in natural photosynthesis,\cite{Scholes:2011kx,Strumpfer:2012ep,Smyth:2012ex}
however, its connection to geometrical factors and EET efficiency
have not been fully investigated. Therefore, in this work we focus
on the discussion of mechanisms related to exploiting exciton delocalization
for efficient EET. 

Regarding the design of efficient EET materials, the spatial arrangements
of chromophores clearly are important variables. However, studies
on the relationship between the spatial arrangement and EET efficiency
in light harvesting have been sporadic. Schulten and coworkers have
emphasized the importance of quantum coherence due to spatial closeness
in affecting the spectral properties and functions of bacterial light-harvesting
apparatus.\cite{Sener:2011p91998,Strumpfer:2012ep} Sun and coworkers
demonstrated that dimerization of bacteriochlorophylls in the ring-like
light-harvesting complexes from purple bacteria leads to an enhanced
efficiency for EET because of the symmetry properties of the eigenstates.\cite{Yang:2010p82080}
Furthermore, Buchleitner and coworkers studied EET efficiency of random
networks of particles using a model with isotropic interactions and
phenomenological incoherent hopping rates to discover properties of
optimal EET networks.\cite{Scholak:2011fv,Scholak:2011dy} Recently,
Scholes and coworkers have investigated mechanisms for efficient EET
between long-distance pairs of pigments in photosynthetic complexes
from marine cryptophytes.\cite{Collini:2010p79587,Marin:2011p99508}
These works have provided much insights into the remarkable efficiency
of natural light-harvesting systems, however, we still do not possess
clear structural design principles for efficient EET systems. 

Natural photosynthetic complexes often exhibit clusters of chromophores
with strong electronic couplings between the cluster members,\cite{Cogdell:2008p59896}
and it has been proposed that these elements play important roles
in the remarkable efficiency of natural light harvesting.\cite{Huo:2011p95856,Pachon:2012fm}
However, the mechanisms responsible for the efficiency enhancement
(if there is any) in a clustered network and the role of electronic
coherence in the process remain unclear. In this work, we aim to discover
spatial arrangements of chromophores that can be used to exploit quantum
coherence for efficient light harvesting. To this end we investigate
a simple linear network of four chromophores. Because of the simplicity,
it allows us to directly elucidate geometrical factors and quantum
coherence effects for the optimization of EET efficiency. We show
that when exciton delocalization is considered, the spatial arrangement
of the sites leading to optimal EET efficiency is different from what
is predicted by a classical theory. As a result we reveal\lyxdeleted{Yuan-Chung Cheng}{Wed Jul 17 17:47:55 2013}{
} new geometrical factors to improve EET efficiency and molecular
design via exciton delocalization. In addition, we argue that (1)
an energy gradient towards the target site and (2) clustered arrangements
of sites that utilize exciton delocalization to tune excitation energies
and open up efficient EET pathways could be considered as two major
factors in building an efficient EET system. Moreover, we show that
these two principles can be applied to explain the design of the electronic
Hamiltonian of the well-studied Fenna-Matthews-Olson (FMO) complex
from green sulfur bacteria. In other words, instead of showing that
the EET in FMO is efficient, we show how FMO is designed to have efficient
EET.

In this work, we focus on electronic quantum coherence induced by
coherent excitonic couplings between chromophores (sites).\cite{Cheng:2009p75665,Scholes:2010p80927,Ishizaki:2012kf,Smyth:2012ex}
It is interesting to note that there are emerging evidences that vibrational
coherence could also play a significant role in photosynthesis.\cite{Kolli:2012ip,Rey:2013ih,Chin:2013ia,Tiwari:dt}
Recently, Plenio and coworkers have proposed a vibrational spectral
tuning principle for efficient EET.\cite{Rey:2013ih} Since their
focus is on the design of environments (bath) while our focus is on
the system's spatial arrangements of chromophores, we believe our
discovery is complementary to their findings. The combined results
might lead to a full set of applicable rules for the molecular design
of the environments (phonons) as well as the system for efficient
EET materials. 

We consider a model system with four chromophores in a linear geometry
as depicted in Scheme \ref{Scheme01}(a). The four sites consist of
a pair of donors and a pair of acceptors determined by their relative
site energies. Since it is well known that a moderate energy gradient
towards the energy sink contributes significantly in natural light-harvesting
systems,\cite{Scholes:2011kx} we choose to consider a case in which
the two donor sites are higher in energy than the two acceptor sites.
Therefore, the linear chain is effectively an energy wire that passes
energy from one end to the other. To investigate the geometrical factors
in EET, we fix the end-to-end distance of the system ($R$) and vary
the intra-pair distance between the two donor (acceptor) sites ($r$).
In addition, we consider orientational dependence ($\theta$) to take
into account the dipole orientation effects. This four-site linear
model, albeit a significant oversimplification, is not only one of
the simplest EET systems that can exhibit electronic coherence effects
but also a faithful representation of highly-ordered molecular arrays
in several synthetic systems.\cite{Huang:2011p93063,Liu:2010p95490}
By studying such a simplified model instead of a natural photosynthetic
complex, we aim to clearly identify key geometrical factors contributing
to efficient EET. 

\begin{scheme}
\includegraphics[width=0.8\columnwidth]{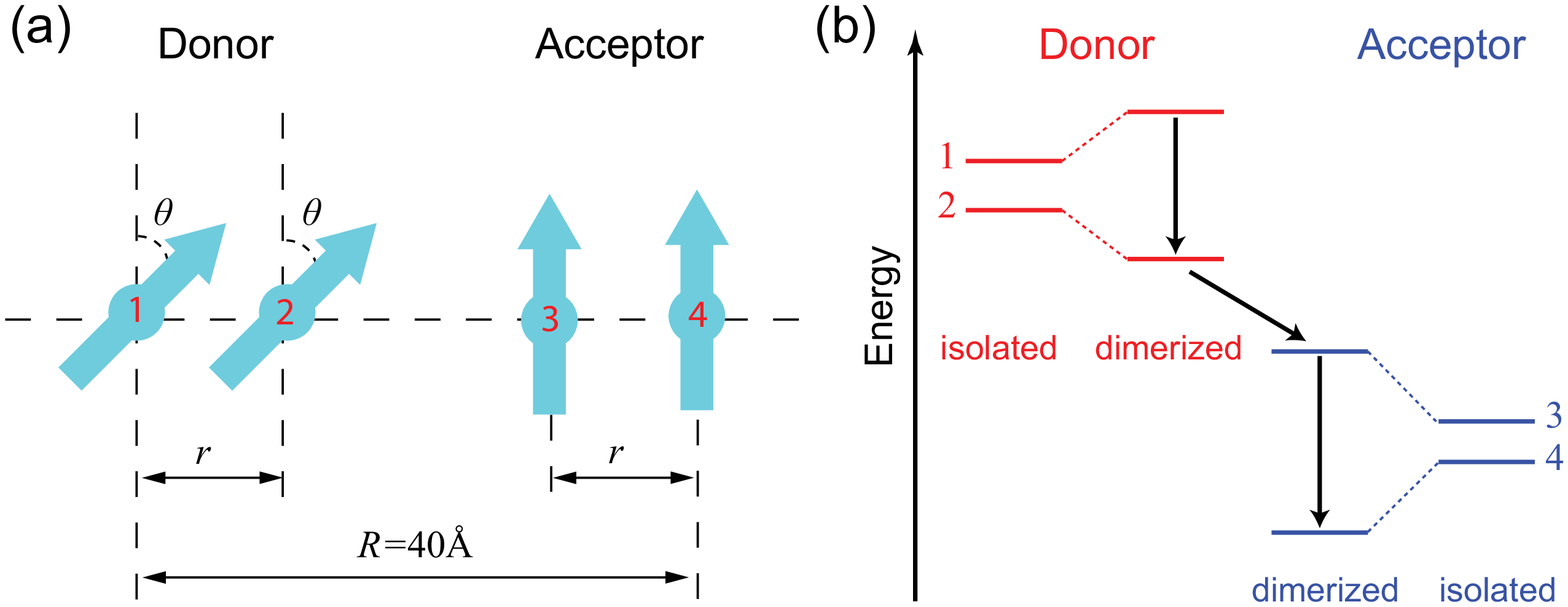}

\caption{\label{Scheme01}(a) Schematic diagram of the four-site linear model
system. In the model, sites 1 and 2 form the donor pair, and sites
3 and 4 are the acceptor pair. The total distance between the two
ends is fixed at $R=40\mathring{A}$, while the intra-pair distance
$(r)$ and dipole tilt angle $(\theta)$ are varied to investigate
geometrical effects. (b) An energy transfer pathway between donor
and acceptor exciton states in the four-site model. The spectral overlap
between the low-energy donor state and high-energy acceptor state
can be enhanced by dimerization, leading to enhanced EET efficiency. }
\end{scheme}

We model the four-site linear system using a Frenkel exciton Hamiltonian:\cite{Cheng:2009p75665,Novoderezhkin:2010p82122}
\begin{equation}
H_{S}=\sum_{j=1}^{4}\varepsilon_{j}\left|j\right\rangle \left\langle j\right|+\sum_{i\neq j}J_{ij}\left|i\right\rangle \left\langle j\right|,
\end{equation}
where $\left|j\right\rangle $ describes the state with single excitation
in the $j$th site, $\varepsilon_{j}$ denotes the site energy of
$\left|j\right\rangle $, and the excitonic coupling $J_{ij}$ is
described by dipole-dipole interactions: 
\begin{equation}
J_{ij}=\frac{1}{4\pi\epsilon_{0}r_{ij}^{3}}\left[\vec{\mu}_{i}\cdot\vec{\mu}_{j}-3\left(\vec{\mu}_{i}\cdot\hat{r}_{ij}\right)\left(\vec{\mu}_{j}\cdot\hat{r}_{ij}\right)\right],
\end{equation}
where $\vec{r}_{ij}=r_{ij}\hat{r}_{ij}$ is the displacement vector
from site $i$ to site $j$, $\vec{\mu}_{i}$ is the transition dipole
of site $i$, and $\epsilon_{0}$ is the vacuum permittivity. In our
numerical simulations, we use the following parameters: $\varepsilon_{1}=13000\textrm{\textrm{cm}}^{-1}$,
$\varepsilon_{2}=12900\textrm{\textrm{cm}}^{-1}$, $\varepsilon_{3}=12300\textrm{\textrm{cm}}^{-1}$,
$\varepsilon_{4}=12200\textrm{\textrm{cm}}^{-1}$, $\mu_{j}=7.75$Debye,
$R=40\mathring{A}$, $r=6\mathring{A}$ to $14\mathring{A}$, and
$T=300$K. To describe system-bath couplings, we consider independent
harmonic baths coupled diagonally to the system described by Ohmic
spectral density functions, as are normally used in the literatures
for photosynthetic complexes (more details about our model system
and parameters are described in Appendix I in the Supporting Information).
Here we remark that \emph{all the parameters chosen in this study
are in accordance with those in natural photosynthetic systems at
the ambient condition}. Therefore, this model provides a realistic
representation to shed light on the physical factors promoting efficient
EET in photosynthetic systems. Note that we have also examined models
with different site energies and intermolecular distances, and the
results are only marginally different and do not affect the conclusions
of this work.

Accurate simulations of the EET dynamics for the model system turn
out to be a great challenge because of the broad parameter range spanned
by varying $r$ and $\theta$. By changing $r$ and $\theta$, the
inter-site couplings ($J_{ij}$s) vary significantly, effectively
turning the system from the strong electronic coupling to weak electronic
coupling limits. Among methods for EET dynamics accurate in a broad
parameter range, the modified-Redfield theory\cite{Zhang:1998p3588,Yang:2002p3160}
yields reliable population dynamics of excitonic systems and has been
successfully applied to describe spectra and population dynamics in
many photosynthetic complexes.\cite{Novoderezhkin:2006p516,Novoderezhkin:2007p7640,Novoderezhkin:2010fb,Novoderezhkin:2010p82122,Novoderezhkin:2011ha}
Following the main idea of the modified-Redfield theory, we consider
the problem in the exciton basis and include the diagonal part of
the exciton-bath interaction operator into the zero-th order Hamiltonian,
while treating only the off-diagonal part of the exciton-bath interaction
operator as the perturbation. We than derive the equations of motion
for the full reduced density matrix of the excitonic system based
on perturbative cumulant expansion up to the second order and the
secular approximation to\lyxdeleted{Yuan-Chung Cheng}{Wed Jul 17 17:47:55 2013}{
} generalize the modified-Redfield theory to treat time evolutions
of off-diagonal terms in a density matrix, \cite{YuHsienHwangFu:2012we}
providing a new theoretical approach for accurate descriptions of
coherent EET dynamics. We thus apply this coherent modified-Redfield
theory (CMRT) to simulate the EET dynamics in this work. Note that
although more accurate nonperturbative methods are available,\cite{Tanimura:2006p3161,Jin:2008p60247,Ishizaki:2009p75287}
they are not applicable in this case due to the need for numerical
efficiency.

Incorporating the time evolution for the off-diagonal terms of the
reduced density matrix in addition to dynamics of diagonal terms governed
by the modified Redfield theory, the CMRT method provides a generalized
quantum master equation (QME) for the reduced density matrix of an
excitonic system in the exciton basis: 
\begin{eqnarray}
\dot{\rho} & = & -i\sum_{kk^{\prime}}(\varepsilon_{k}^{\prime}-\varepsilon_{k^{\prime}}^{\prime})\rho_{kk^{\prime}}\left\vert \varepsilon_{k}\right\rangle \left\langle \varepsilon_{k^{\prime}}\right\vert +\sum_{k\neq k^{\prime}}\left(R_{kk^{\prime}}^{\textrm{dis}}(t)\rho_{k^{\prime}k^{\prime}}\left\vert \varepsilon_{k^{\prime}}\right\rangle \left\langle \varepsilon_{k^{\prime}}\right\vert -R_{k^{\prime}k}^{\textrm{dis}}(t)\rho_{kk}\left\vert \varepsilon_{k}\right\rangle \left\langle \varepsilon_{k}\right\vert \right)\nonumber \\
 &  & -\sum_{k\neq k^{\prime}}\left[R_{kk^{\prime}}^{\textrm{pd}}(t)+\frac{1}{2}\sum_{k^{\prime\prime}\left(\neq k,k^{\prime}\right)}\left(R_{k^{\prime\prime}k}^{\textrm{dis}}(t)+R_{k^{\prime\prime}k^{\prime}}^{\textrm{dis}}(t)\right)\right]\rho_{kk^{\prime}}\left\vert \varepsilon_{k}\right\rangle \left\langle \varepsilon_{k^{\prime}}\right\vert ,\label{eqOriCMRT}
\end{eqnarray}
where $\rho(t)$ is the density matrix with matrix element $\rho_{kk^{\prime}}=\left\langle \varepsilon_{k}\right\vert \rho\left\vert \varepsilon_{k^{\prime}}\right\rangle $,
$\left\vert \varepsilon_{k}\right\rangle $ is the $k$th eigen state
of the electronic Hamiltonian $H_{S}$, and $\varepsilon_{k}^{\prime}$
is the eigen energy of $\left\vert \varepsilon_{k}\right\rangle $
shifted by a reorganization energy due to the system-bath coupling.
The expressions for time-dependent dissipation rates ($R_{kk^{\prime}}^{\textrm{dis}}(t)$)
and pure-dephasing rates ($R_{kk^{\prime}}^{\textrm{pd}}(t)$) are
given in Appendix II in the Supporting Information. The CMRT has clear
physical interpretation for each terms. The three terms describe the
coherent dynamics, the population dynamics, and the dephasing processes
due to pure dephasing and population-transfer induced decoherence,
respectively. In addition, since we consider time-dependent rates,
the CMRT QME is non-Markovian.

Furthermore, an efficient numerical simulation scheme is required
as we need to calculate long-time dynamics of the four-site model
with many different $r$ and $\theta$ in order to evaluate geometrical
effects in the efficiency of light harvesting. To this end we have
adopted a non-Markovian quantum trajectory (NMQT) method\cite{Piilo:2008di,Piilo:2009p89261}
to provide efficient simulations of the CMRT dynamics. Specifically,
direct propagation of the CMRT QME for a system with $M$ sites requires
solving $M^{2}$ ordinary differential equations, whereas the NMQT
method enables more efficient simulations of EET dynamics by unraveling
the non-Markovian QME into $M$ stochastic Schr\"{o}dinger equations.
Details on the unraveling of the CMRT QME and numerical algorithms
for time propagation are given in Appendix III in the Supporting Information.
Since the accuracy of the modified-Redfield theory is well validated
for biased-energy systems,\cite{Novoderezhkin:2010p82122} the combined
CMRT-NMQT approach provides an accurate and numerically efficient
means to calculate coherent EET dynamics for general multi-level excitonic
systems. Moreover, we have found that the NMQT simulation of the CMRT
QME delivers additional numerical stability that is required to provide
information about long-time EET dynamics governing the efficiency
of EET.

\begin{figure}
\includegraphics[width=1\columnwidth]{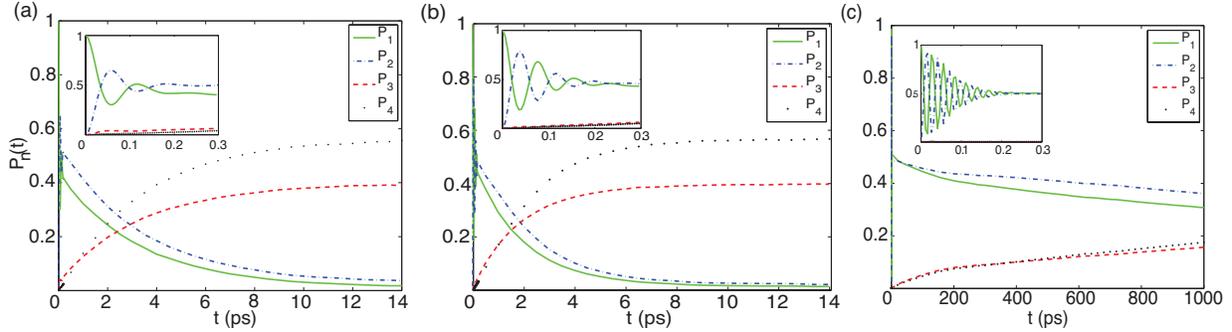}\caption{\label{fig:Excitation-energy-transfer}The EET dynamics of the four-site
model calculated by the CMRT-NMQT method. The curves show populations
of excitation on each of the four molecules in the system. (a) population
dynamics for the geometry $r=13.4\mathring{A}$ and $\theta=0$, (b)
population dynamics for the geometry $r=11.4\mathring{A}$ and $\theta=0$,
(c) population dynamics for the geometry $r=8\mathring{A}$ and $\theta=0$.
The insets display the coherent evolutions at the short times. }
\end{figure}

We applied the combined CMRT-NMQT approach to investigate EET dynamics
in the four-site model. To mimic the function of the system as a wire
transferring energy from site 1 to site 4, we assume site 1 is initially
excited in all our calculations. In Fig. \ref{fig:Excitation-energy-transfer}
we present calculated site population dynamics for the four-site model
in three geometries with different intra-pair distances. Figure \ref{fig:Excitation-energy-transfer}(a)
shows the EET dynamics of a system with equally-spaced sites, i.e.,
$r=13.4\mathring{A}$. In this case the excitation energy moves coherently
between sites 1 and 2 in $\sim300$ fs and then is transferred from
the donor to the acceptor in $\sim4$ ps timescale. Clearly, the combination
of coherent energy relaxation within the donor pair and incoherent
EET from the donor pair to the acceptor pair governs the whole EET
process. Notably, this feature also appears in a broad range of models
for natural light-harvesting systems.\cite{Scholes:2011kx,Pachon:2012fm}
Furthermore, we note that the combined CMRT-NMQT approach allows us
to efficiently capture both the coherent dynamics at short times and
the population decay at long times, while the original modified-Redfield
approach ignores the coherent dynamics and as a result it can not
be applied to describe the dynamics investigated here.

Figure \ref{fig:Excitation-energy-transfer}(b) shows the EET dynamics
of the four-site model with a reduced intra-pair distance ($r=11.4\mathring{A}$).
Compared to Fig. \ref{fig:Excitation-energy-transfer}(a), EET from
donor to acceptor in this dimerized geometry is significantly more
rapid ($\sim2$ ps). This result shows that the EET dynamics are sensitive
to the geometry of the system, and the equally-spaced geometry (Fig.
\ref{fig:Excitation-energy-transfer}(a)) does not necessarily offer
the best EET efficiency. Moreover, as the intra-pair distance $r$
is further decreased to a smaller value ($r=8\mathring{A}$, Fig.
\ref{fig:Excitation-energy-transfer}(c)), the EET becomes extremely
slow ($\sim7$ ns). Despite this, the oscillations in the short-time
regime remain significant and even more pronounced due to stronger
intra-pair electronic coupling strengths. The results presented in
Fig. \ref{fig:Excitation-energy-transfer} indicate that an optimal
geometry for EET exists, and the efficiency of EET is closely related
to quantum coherence but not always positively correlated with the
coherent evolution. 

To quantify the efficiency of energy transfer across the four-site
model system, we study the population at site 4 as a function of time,
$P_{4}(t)$. As shown in Fig. \ref{fig:Excitation-energy-transfer},
$P_{4}(t)$ generally exhibits a double-exponential behavior, with
a rapid but minor rise and a major long-time exponential growth. Therefore,
we fit $P_{4}(t)$ to a double-exponential growth model and take the
major rise component to define an effective end-to-end transfer rate
$R_{\textrm{eff}}$ (see Appendix IV in the Supporting Information).
For the geometries studied in Fig. \ref{fig:Excitation-energy-transfer},
the effective transfer times $\tau_{\textrm{eff}}=1/R_{\textrm{eff}}$
obtained by fitting to $P_{4}(t)$ are respectively: (a) $\tau_{\textrm{eff}}=3.4$ps,
(b) $\tau_{\textrm{eff}}=2.4$ps, and (c) $\tau_{\textrm{eff}}=6.9$ns. 

\begin{figure}
\includegraphics[width=0.6\columnwidth]{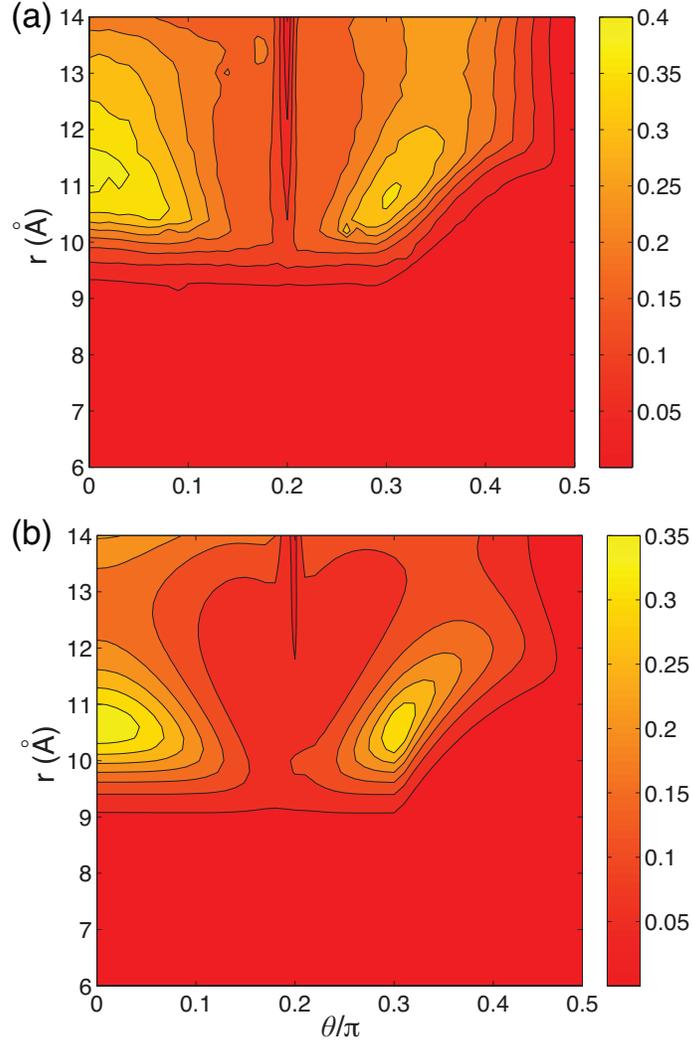}\caption{\label{fig:Effective-transfer-rate}Geometrical dependence of the
effective transfer rate $R_{\textrm{eff}}$ (ps$^{-1}$) as a function
of the two geometrical parameters $r$ and $\theta$. (a) $R_{\textrm{eff}}$
calculated from full CMRT-NMQT dynamical simulations. (b) $R'_{\textrm{eff}}$
predicted by the analytical result in Eq. \ref{eq:Reff2R}. }
\end{figure}

In order to further investigate the relation between the EET efficiency
and spatial arrangement of the sites, we simulate the EET dynamics
of the four-site model for a broad range of intra-pair distance $r$
and dipole orientation $\theta$ to obtain the effective end-to-end
transfer rates $R_{\textrm{eff}}$. Figure \ref{fig:Effective-transfer-rate}(a)
shows $R_{\textrm{eff}}$ as a function of $r$ and $\theta$. This
map shows two optimal regions for the efficiency of EET across the
linear four-site model near $\left(\theta=0,r=11.3\mathring{A}\right)$
and $\left(\theta=0.3\pi,r=10.8\mathring{A}\right)$, of which the
intra-pair distances are clearly smaller than the equally-spaced value
($r=13.4\mathring{A}$). This indicates that dimerized clusters offer
enhanced transfer rates. For the optimal region at $\theta=0$, the
effective transfer rate increases to a maximum value as the distance
is decreased to $r=11.3\mathring{A}$ from the equally-spaced value.
As the distance is further decreased, the rate of energy transfer
quickly decreases. This suggests that EET between the donor pair and
the acceptor pair becomes crucial as the inter-pair distance becomes
large. 

The enhanced EET efficiency at dimerized geometries can be attributed
to improved energy matching between donor and acceptor states due
to increased intra-pair electronic couplings. As illustrated in Scheme
\ref{Scheme01}(b), the exciton delocalization in the donor and acceptor
pairs can shift eigen-energies and reduce the energy gap between the
lower-energy donor state and higher-energy acceptor state. When combined
with rapid coherent equilibration of energy among the donor states,
the enhanced energy matching leads to more efficient EET from the
donor to the acceptor. Note that a classical sequential hopping picture
of EET such as the Förster theory predicts that the equally-spaced
geometry would give rise to the optimal end-to-end EET efficiency
because otherwise the slowest step would limit the transfer rate.
What we have revealed here is an intrinsically quantum mechanical
mechanism based on energy tuning due to exciton delocalization that
gives rise to enhanced EET efficiency of a non-trivial dimerized geometry.
We remark that the precise optimal geometry for EET depends on how
the electronic couplings between sites are calculated. In our model,
we assume dipole-dipole interactions, which may not be valid in the
short-distance regime. Nevertheless, more accurate estimations of
electronic couplings should not change the prediction that dimerized
clusters will give rise to enhanced EET efficiency in this particular
geometry.

\begin{figure}
\includegraphics[width=0.6\columnwidth]{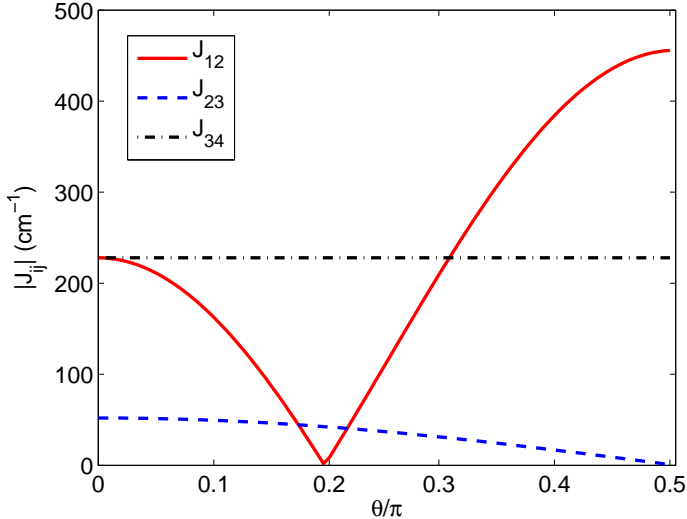}\caption{\label{fig:JvsTheta}Magnitude of the calculated electronic coupling
strengths $\left|J_{ij}\right|$ as a function of $\theta$ for $r=10.8\mathring{A}$.
Here we plot the three nearest-neighbor couplings as other couplings
are much smaller. Note that because $J_{12}<0$ for $\theta>0.2\pi$,
the donor pair forms an H-aggregate when $\theta<0.2\pi$, and becomes
a J-aggregate for $\theta>0.2\pi$.}
\end{figure}

Regarding the dependence of $R_{\textrm{eff}}$ on $\theta$, the
situation becomes more complicated. In addition to the region around
$\theta=0$ (parallel transition dipoles), there exists the other
optimal region around $\theta=0.3\pi$ in Fig. \ref{fig:Effective-transfer-rate}(a).
These two regions are separated by a minimum $R_{\textrm{eff}}$ around
$\theta\simeq0.2\pi$. To elucidate the orientational effects, we
plot the magnitudes of the nearest-neighbor electronic couplings,
$J_{12}$, $J_{23}$, and $J_{34}$, as a function of $\theta$ in
Fig. \ref{fig:JvsTheta}. There, $J_{34}$ remains constant because
the orientations of the two acceptor dipoles are fixed. In contrast,
$J{}_{23}$ slowly and monotonically decreases to zero as $\theta$
increases from $0$ to $0.5\pi$, while the dipoles of sites $2$
and $3$ change from a parallel orientation to a perpendicular orientation.
Moreover, $J_{12}$ not only changes its magnitude significantly but
also its sign at $\theta\simeq0.2\pi$, giving rise to a negligible
$J_{12}$ around $\theta=0.2\pi$. As a result of the vanishing $J_{12}$,
energy transfer is strongly suppressed at $\theta\simeq0.2\pi$, leading
to the local minimum of $R_{\mathrm{eff}}$ in Fig. \ref{fig:Effective-transfer-rate}(a).
Clearly, the optimal region at $\theta=0.3\pi$ is an interplay of
the increase in the donor intra-pair coupling ($J_{12}$) and decrease
in the inter-pair coupling ($J_{23}$) as $\theta$ passes $0.2\pi$.
Finally, it is intriguing to note that the redistribution of transition
dipoles due to electronic coherence also plays a minor role in the
optimal region at $\theta=0.3\pi$, since the geometry corresponds
to EET from a J-aggregate dimer to an H-aggregate dimer, indicating
that enhanced transition dipole in the lower-energy state of the donor
(J-type) and the higher-energy state of the acceptor (H-type) could
further enhance the rate of donor-to-acceptor EET.

Figure \ref{fig:Effective-transfer-rate}(a) maps out the geometrical
control factors of the effective transfer rate for the four-site model.
What is the EET mechanism responsible for the geometrical dependence
of EET efficiency? Does coherent evolution play a significant role
in the optimal geometries? Can we establish a simple EET picture that
accurately describes the dynamics in the broad parameter range covered
in Fig. \ref{fig:Effective-transfer-rate}(a)? To answer these questions
and elucidate the key mechanisms of EET in the four-site model, we
consider EET in the system as a cascade energy relaxation involving
three sub-processes (Scheme \ref{Scheme01}(b)): $\left\vert \varepsilon_{1}\right\rangle \rightarrow\left\vert \varepsilon_{2}\right\rangle $,
$\left\vert \varepsilon_{2}\right\rangle \rightarrow\left\vert \varepsilon_{3}\right\rangle $,
and $\left\vert \varepsilon_{3}\right\rangle \rightarrow\left\vert \varepsilon_{4}\right\rangle $. 

Due to the strong electronic coupling between the donor sites, there
are significant populations distributed on both $\left\vert \varepsilon_{1}\right\rangle $
and $\left\vert \varepsilon_{2}\right\rangle $ when site 1 is initially
excited. Therefore, two dominant energy relaxation pathways should
be considered: $\left\vert \varepsilon_{1}\right\rangle \rightarrow\left\vert \varepsilon_{2}\right\rangle \rightarrow\left\vert \varepsilon_{3}\right\rangle $
and $\left\vert \varepsilon_{2}\right\rangle \rightarrow\left\vert \varepsilon_{3}\right\rangle \rightarrow\left\vert \varepsilon_{4}\right\rangle $.
The two relaxation processes compete with each other and the overall
transfer time is determined by the slower one that is rate-determining.
Consequently, if we assume the transfer proceed through exciton states
in a sequential manner in the exciton basis, then the time required
to complete the overall transfer process can be calculated from the
rates of each individual steps: 
\begin{eqnarray}
\tau_{\textrm{total}} & \simeq & \max\left\{ \tau_{2\leftarrow1}+\tau_{3\leftarrow2},\tau_{3\leftarrow2}+\tau_{4\leftarrow3}\right\} \nonumber \\
 & \simeq & \max\left\{ \frac{1}{R_{21}^{\prime}}+\frac{1}{R_{32}^{\prime}},\frac{1}{R_{32}^{\prime}}+\frac{1}{R_{43}^{\prime}}\right\} \simeq\frac{1}{R_{\textrm{eff}}^{\prime}},
\end{eqnarray}
where $R_{21}^{\prime}$, $R_{32}^{\prime}$, and $R_{43}^{\prime}$
correspond to state-to-state EET rates of $\left\vert \varepsilon_{1}\right\rangle \rightarrow\left\vert \varepsilon_{2}\right\rangle $,
$\left\vert \varepsilon_{2}\right\rangle \rightarrow\left\vert \varepsilon_{3}\right\rangle $,
and $\left\vert \varepsilon_{3}\right\rangle \rightarrow\left\vert \varepsilon_{4}\right\rangle $,
respectively. As a result, we can estimate the effective transfer
rate as 
\begin{eqnarray}
R_{\textrm{eff}}^{\prime} & \equiv & \min\left\{ \frac{R_{21}^{\prime}R_{32}^{\prime}}{R_{21}^{\prime}+R_{32}^{\prime}},\frac{R_{32}^{\prime}R_{43}^{\prime}}{R_{32}^{\prime}+R_{43}^{\prime}}\right\} .\label{eq:Reff2R}
\end{eqnarray}
To calculate the three relevant EET rates, $R_{21}^{\prime}$, $R_{32}^{\prime}$,
and $R_{43}^{\prime}$, we opt for a combined modified-Redfield and
generalized Förster approach.\cite{Novoderezhkin:2011ha} Because
of the relatively-strong dipole-dipole interactions within the donor
and acceptor pairs, we calculate intra-pair transfer rates $R_{21}^{\prime}$
and $R_{43}^{\prime}$ using the CMRT approach. On the other hand,
due to the relatively weak inter-pair dipole-dipole interactions,
the transfer rate from the donor to acceptor, $R_{32}^{\prime}$,
is calculated by the multi-chromophoric Förster resonance energy transfer
(MC-FRET) theory.\cite{Jang:2004ica} As a result, we describe the
EET processes in the dimerized four-site system as rapid relaxation
within the respective donor and acceptor pairs described by CMRT and
slower incoherent hopping of exciton from donor to acceptor described
by the MC-FRET. We then use the calculated EET rates to estimate the
effective transfer rate by Eq. (\ref{eq:Reff2R}) (see Appendix V
in the Supporting Information). 

The estimated rates are presented in Fig. \ref{fig:Effective-transfer-rate}(b).
The excellent agreement of the estimated results to Fig. \ref{fig:Effective-transfer-rate}(a),
which is obtained from full dynamical simulations, indicates that
the combined CMRT/MC-FRET mechanism provides an excellent picture
to describe the EET dynamics in the four-site model. We thus arrive
at the conclusion that instead of the time-consuming full simulation
of quantum dynamics, we could efficiently estimate the effective efficiency
of the complete EET process by the combined CMRT/MC-FRET picture.
Furthermore, in the analysis based on the combined CMRT/MC-FRET picture,
eigenstates of the donor pair are used as the initial states for calculating
the overall transfer rate determined by the slower process in the
two-pathway model (Appendix V in the Supporting Information). Therefore,
the excellent agreement between Fig. \ref{fig:Effective-transfer-rate}(a)
and Fig. \ref{fig:Effective-transfer-rate}(b) also indicates that
although a localized initial state is used for propagating population
dynamics by the combined CMRT-NMQT approach, the optimal geometries
do not depend on such an initial preparation. In addition, since the
estimated transfer rate is calculated from a purely incoherent picture
of energy relaxation in the delocalized exciton basis, the result
indicates that coherence evolution does not play a significant role
in the optimization of EET efficiency in this system. Note that this
is likely related to the model system chosen here because coherent
dynamics only contribute to rapid equilibrium within the donor pair
and have negligible effects on the donor-to-acceptor EET process probed
by the effective transfer rate. The coherently wired EET between long-distance
pairs of pigments observed in photosynthetic complexes from marine
cryptophytes\cite{Collini:2010p79587,Marin:2011p99508} shows that
the coherent dynamics could play an additional role in the optimization
of EET efficiency. However, in this work we only investigate how to
use geometrical factors to exploit exciton delocalization for enhanced
EET efficiency. It will be interesting to investigate geometries that
can take advantage of coherent evolution in the future work.

Finally, what are the geometrical factors governing efficient EET
in natural photosynthetic systems? Are the coherence-assisted principles
revealed here applicable to the chlorophyll arrangements in photosynthetic
complexes? Our calculations based on the four-site model suggest a
rapid EET can be achieved by (1) employing an energy bias towards
the target site to create directional energy flow and (2) forming
strongly coupled clusters to create exciton delocalization and tune
exciton energies for better energy matching between the donor and
acceptor cluster states. It is a common observation that natural photosynthetic
complexes often exhibit moderate energy bias and clusters of chromophores,\cite{Huo:2011p95856,Scholes:2011kx,Pachon:2012fm}
which are fully consistent with our results. Moreover, the coherence-assisted
energy tuning and energy bias clearly play crucial roles in the rapid
EET in bacterial light-harvesting systems\cite{Sener:2011p91998,Strumpfer:2012ep,Jang:2012fi}
and the long-range coherent energy transfer in cryptophytes.\cite{Novoderezhkin:2010fb,Huo:2011p95856}
In particular, the dimerization in the B850 ring of the light-harvesting
system 2 (LH2) from purple bacteria leads to the spreading of the
energy levels of the B850 states, which enables rapid energy transfer
from B800 to upper B850 states.\cite{Yen:2011p99898} In the following
we apply the two principles to shed light on the construction of the
effective exciton Hamiltonian of the FMO complex from green sulfur
bacteria.

\begin{figure}
\includegraphics[width=0.9\columnwidth]{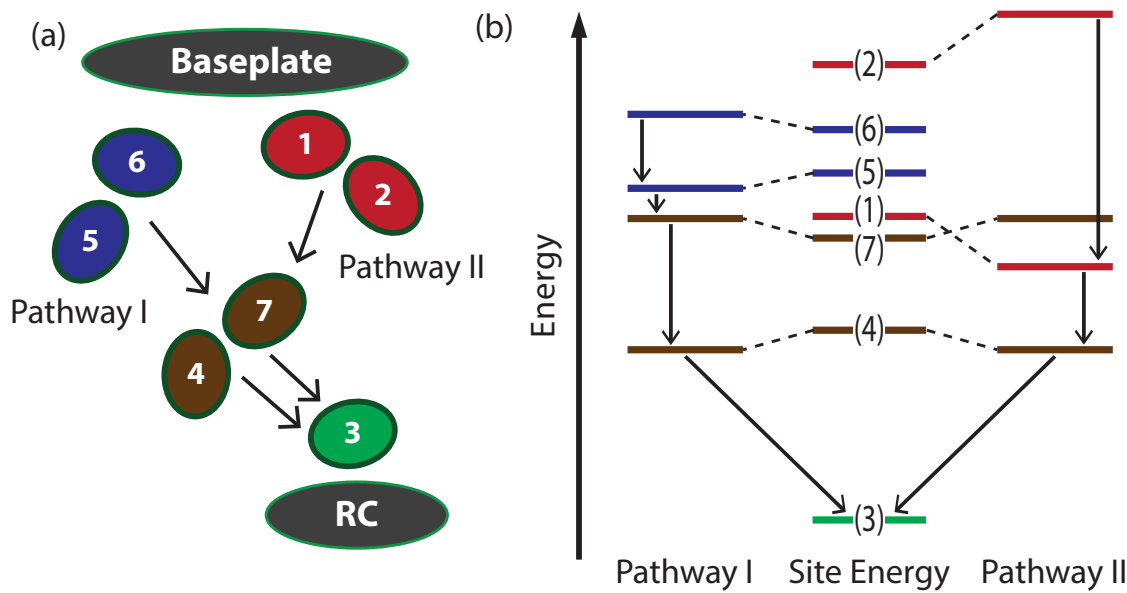}

\caption{\label{fig:FMO}(a) Chlorophyll arrangement and EET pathways in the
monomeric subunit of the FMO complex with respect to the baseplate
and reaction center. (b) Excitation energy of each chlorophyll in
FMO and the energetics of the exciton states.\cite{Cho:2005p4861}
Each of the two EET pathways is composed of a donor dimer and an acceptor
dimer, similar to the four-site model studied in this work. }
\end{figure}

The FMO complex exhibits remarkable energy transfer efficiency that
has become the subject of intensive research.\cite{Brixner:2005wu,Cho:2005p4861,Ishizaki:2009p77158,Wu:2010bg,Moix:2011gi}
Figure \ref{fig:FMO}(a) shows the pigment arrangement in a monomeric
subunit of the FMO complex. FMO contains 7 bacteriochlorophylls (BChls),
and its function is to conduct energy from the baseplate (close to
BChls 1 and 6) to the reaction center (close to BChl 3). Therefore,
it is conceivable that an energy gradient should exist to direct the
energy flow from one end of the complex to the other end. Indeed,
effective Hamiltonian models that were obtained from spectral fitting\cite{Cho:2005p4861,Adolphs:2006p4869}
and quantum chemistry calculations\cite{Mueh:2007p8876} all indicate
that BChl 3 has the lowest transition energy, BChls 1, 2, 5, and 6
have higher energies, and the energies of BChls 4 and 7 are set in
between (Fig. \ref{fig:FMO}(b)). Clearly, the site energies in FMO
are tuned to form an energy funnel towards the reaction center. Noticeably,
the energy of BChl 3 is significantly lower than all other sites.
The large energy gap could support a highly localized exciton state
and prevent back transfer, effectively generating a higher thermal
population of excitation energy on BChl 3 for more efficiency EET
to the reaction center. Thus, the site energies of BChls in FMO are
fully consistent with the energy bias principle for enhancing EET
efficiency.

In addition to site energies, electronic couplings can be varied by
geometrical dimerization to improve energy matching of exciton states
for efficient EET, and this is exactly the case in FMO. The Pathway
I shown in Fig. \ref{fig:FMO}(b) clearly illustrates the similarity
to the dimerization scheme shown in Scheme \ref{Scheme01}(b). There,
BChls 5 and 6 are dimerized, and so are BChls 4 and 7, leading to
large intra-pair electronic couplings (i.e. large $J_{56}$ and $J_{47}$)
and delocalized excitations. As a result of the exciton delocalization
caused by the enhanced couplings, energy gap between the lower-energy
exciton state of the BChls 5-6 dimer (donor) and the higher-energy
exciton state of the BChls 4-7 dimer (acceptor) is reduced, which
leads to increased spectral overlap and enhanced EET rate from the
donor to the acceptor according to MC-FRET. For the intra-pair transfer,
although the energy gap between two exciton states is increased, the
energy transfer can still be rapid because intra-pair energy relaxation
between delocalized states follow the CMRT mechanism, which mainly
depends on the spatial overlap of exciton states.\cite{Cheng:2009p75665,Novoderezhkin:2010p82122}
Similar mechanism can be used to describe Pathway II, explaining the
two primary EET pathways as have been revealed by 2D electronic spectroscopy.\cite{Brixner:2005wu,Cho:2005p4861}
We argue that energy bias and exciton delocalization could account
for the $>90\%$ quantum efficiency of FMO, while tuning of vibrational
environment (spectral density) and coherent evolution may provide
efficiency optimization.\cite{Cao:2009p78157,Ishizaki:2009p77158,Wu:2010bg,Ishizaki:2012kf,Chin:2013ia,Rey:2013ih,Wu:2013ig}
In other words, we have made clear two key underlying principles for
the design of the effective Hamiltonian of FMO, and the control through
spatial arrangements of BChls is apparent.

In summary, we have investigated the optimal spatial arrangements
of chromophores for EET in a linear four-site model mimicking a coherent
light-harvesting system. Based on combined CMRT-NMQT numerical simulations
of EET dynamics, the geometrical factors affecting the end-to-end
EET rate are investigated, and it is discovered that the effective
transfer rate is maximized if the donor and acceptor sites are respectively
dimerized in this given topology. Moreover, we also demonstrated that
the dipole orientation angle $\theta$ also plays an important role
in the EET efficiency. This result is interesting and non-trivial
because a classical approach such as the Förster resonance energy
transfer theory would predict that an equally spaced linear geometry
should give rise to optimal end-to-end transfer rate. Our analysis
reveals that in contrast exciton delocalization contributes crucially
to the optimization of energy transfer efficiency. We conclude that
coherence-assisted energy tuning based on geometrical control of inter-chromophore
electronic couplings provides a useful means to enhance EET efficiency.
Furthermore, it is already well known that a biased energy funnel
to direct energy flow towards the target site\lyxdeleted{Yuan-Chung Cheng}{Wed Jul 17 17:47:55 2013}{
} also plays an important role for constructing efficient light-harvesting
systems.\cite{Scholes:2011kx} Based on these rules, the Hamiltonian
of FMO is analyzed, and it is shown that the spatial arrangement of
the 7 BChls in combination with the site energy gradient favor efficient
EET from the baseplate to the reaction center. In addition, we argue
that FMO also makes use of exciton delocalization and dimerized BChl
arrangements to optimize energy transfer. Although we have only investigated
a simple linear system with small geometrical degrees of freedom,
these principles could be generalized to larger systems.\lyxdeleted{Yuan-Chung Cheng}{Wed Jul 17 17:47:55 2013}{
} In Appendix VI in the Supporting Information, energy transfer in
a ring-shape six-site model is investigated to demonstrate that without
an exception, clustered geometries also lead to optimal EET in such
a non-linear topology. Therefore, we believe the basic principles
revealed in this work may be generalized to larger molecular networks
and benefit the future innovation of efficient artificial light-harvesting
materials. Note that homogeneous systems such as chlorosome or B850
rings of LH2 from purple bacteria belong to a different class of systems,
in which additional symmetry-related rules might play more prominent
roles in the optimization of EET rate.\cite{Yang:2010p82080,Kim:2010p93059,Cleary:2013dl}

Finally, because we focus on the long-time dynamics in this work,
the short-time coherent dynamics are overlooked, and an in depth investigation
of the coherent evolution effects and geometrical factors might provide
additional optimization rules for light harvesting.\cite{Scholes:2011kx,Huo:2011p95856}
In this regard we believe the combined CMRT-NMQT approach should provide
a powerful theoretical tool for seeking optimal design of both natural
and artificial light-harvesting systems, since it yields accurate
and efficient numerical simulations of coherent quantum dynamics for
large excitonic systems.

\begin{suppinfo}
Details of the effective Hamiltonian and numerical parameters of the four-site model, brief introduction to the CMRT approach, description and numerical implementation of the CMRT-NMQT method, definition of the effective transfer rate,  details of the two-pathway model for EET efficiency, and full calculation and analysis on the EET of a ring-shape siz-site model.
\end{suppinfo}

\begin{acknowledgement}
We thank stimulating discussions with J. Piilo, Y.-X. Hwangfu, Y.-J. Fan, and Y.-H. Liu. QA thanks the National Science Council, Taiwan (Grant No. NSC 100-2811-M-002-162) for financial support. YCC thanks the National Science Council, Taiwan (Grant No. NSC 100-2113-M-002-004-MY2), National Taiwan University (Grant No. 101R891305), and Center for Quantum Science and Engineering (Subproject: 101R891401) for financial support. We are grateful to Computer and Information Networking Center, National Taiwan University for the support of high-performance computing facilities. 
\end{acknowledgement}

\providecommand*\mcitethebibliography{\thebibliography}
\csname @ifundefined\endcsname{endmcitethebibliography}
  {\let\endmcitethebibliography\endthebibliography}{}

\end{document}